\documentclass[
*aip,
reprint,
runinaddress,
showpacs,preprintnumbers,
 amsmath,amssymb,
 aps,
pra,
floatfix,
]{revtex4-1}
\usepackage{mathtools}
\usepackage{graphicx}% Include figure files
\usepackage{dcolumn}% Align table columns on decimal point
\usepackage{bm}% bold math
\usepackage{hhline}
\usepackage{color,soul}
\usepackage{float}
\usepackage{mathrsfs}

\usepackage[normalem]{ulem}
\usepackage[dvipsnames]{xcolor}
\usepackage[sort&compress]{natbib}
\usepackage{xr-hyper}
\usepackage{hyperref}
\usepackage{float}
\hypersetup{
    colorlinks,
    linkcolor={black},
    citecolor={black},
    urlcolor={black}
	}

\newcommand{\MoS}{$\rm{MoS_2}$}
\newcommand{\sio}{$\rm{SiO_2/Si}$}
\newcommand{\cm}{$\rm{cm^{-1}}$}
\newcommand{\ei}{$E_{\rm inc}$}

\begin{document}
\title{Neutral atom scattering based mapping of atomically thin layers}

\author{Geetika Bhardwaj}%
\affiliation{Tata Institute of Fundamental Research Hyderabad, 36/P Gopanpally, Hyderabad 500046, Telangana, India}

\author{Krishna Rani Sahoo}%
\affiliation{Tata Institute of Fundamental Research Hyderabad, 36/P Gopanpally, Hyderabad 500046, Telangana, India}

\author{Rahul Sharma}%
\affiliation{Tata Institute of Fundamental Research Hyderabad, 36/P Gopanpally, Hyderabad 500046, Telangana, India}%

\author{Parswa Nath}%
\affiliation{Tata Institute of Fundamental Research Hyderabad, 36/P Gopanpally, Hyderabad 500046, Telangana, India}%

\author{Pranav R. Shirhatti}
\email[Author to whom correspondence should be addressed. \\ e-mail:{\ }]{pranavrs@tifrh.res.in}
\affiliation{Tata Institute of Fundamental Research Hyderabad, 36/P Gopanpally, Hyderabad 500046, Telangana, India}%

\begin{abstract}

\textbf{Abstract:} Imaging surfaces using low energy neutral atom scattering is a relatively recent development in the field of microscopy.
In this work we demonstrate that this technique is sensitive enough to distinguish films as thin as a single monolayer from the underlying substrate.
Using collimated beams of He and Kr atoms as an incident probe on \MoS{} films grown on \sio{} substrate, we observe systematic changes in the scattered atom flux which allows us to map the thin \MoS{} films.  
Measurements carried out by varying incidence energy using both He and Kr provides insights into the details of atom-surface collision dynamics and its role in contrast generation. 
\end{abstract}

\maketitle

\section{\label{sec:level1}Introduction}
Developments in microscopy and imaging techniques have made a profound impact on our understanding of a wide range of physical phenomena. The variety of techniques developed over the years is impressive – such as the light microscope \cite{Murphy}, sub-$\mathrm{\AA}$ level resolution electron microscopes \cite{ernst,gao2019real}, surface probe techniques such as atomic force and scanning tunneling microscopes \cite{rugar1990atomic, eaton2010atomic,binnig1982surface, binnig1983scanning}, to name a few. 
A relatively recent addition to this gamut of microscopy techniques is Neutral Atom Microscopy (NAM) / Scanning Helium Microscopy (SHeM), when neutral atoms / He atoms are used as an incident probe \cite{allison_2003, witham, barr_RSI, witham_NAM_2014}. 
Its foundations lie in the scattering of low energy neutral atoms from surfaces, an area pioneered by Stern and co-workers many decades ago \cite{estermann_Stern1930}.

Unique features of NAM as an imaging technique are that it is  non-destructive and universally applicable. 
Typical kinetic energy of incident neutral atoms used as probe range from ten to a few hundred meV. Being at least an order of magnitude smaller than chemical bond energies ($\sim$ few\,eV), it is  particularly well suited for imaging delicate surface structures, otherwise susceptible to beam-induced damage.

Given that de Broglie wavelength of atoms at kinetic energies in the above mentioned range is of the order of $\mathrm{\AA}$, in principle, a very high resolution imaging is possible using this approach.
In practice though, the achievable resolution is primarily constrained by the limited ability to manipulate the motion of incident neutral atoms. 
Nonetheless, improvements in detector design \cite{Ward_He_Detector2021} and  experimental schemes to manipulate the incident atoms \cite{holst_theoretical_pinhole,holst_theoretical_model_zoneplate,holst_mirror} hold enormous promise in achieving spatial resolution, well into sub - 100\,nm range and beyond.

Another fundamental aspect of NAM that needs to be looked into further  to realize its full potential i.e. to understand the contrast generation mechanisms leading to image formation. Contrast ($C$) is a measure of the ability to distinguish different features observed in an image, and is defined as the normalized difference in signals $I_1$ and $I_2$ observed from two different surface features \cite{allison_2003} as shown below:
\begin{equation}
C = \left| \frac{I_1 - I_2}{I_1 + I_2} \right|
\label{equation:contrast}
\end{equation}

One of the most commonly observed features in NAM images is that of topographical contrast. This results from a change in the detected signal, caused by changes in diffuse scattering due to different geometric features on the surface \cite{allison_2003}. 
Contrast arising from the changes in scattering distributions resulting from inelastic collisions has also been identified \cite{barr_NatComm2016}. 
More recently, contrast resulting via atomic diffraction by the crystalline nature of target sample has also been demonstrated \cite{jardine_diffraction}. 
This opens up the possibility of seeing features beyond the capability of conventional imaging methods.
It is well known that atom scattering from surfaces is sensitive to local surface structure and disorder at an atomic length scale \cite{farias_rev1998} and in principle can be utilized for contrast generation. However, it is generally expected that in order to observe these effects, precisely prepared clean surfaces under ultra-high vacuum conditions are needed.

In this work, we have studied atom scattering based imaging of thin \MoS{} films grown on \sio{} substrate. 
We take advantage of the fact that such two-dimensional (2D) materials are relatively chemically inert and need less stringent working conditions \cite{farias_graphenemirror_2011}.
Our results demonstrate that \MoS{} films, as thin as a single monolayer, can be distinguished from the substrate and mapped by means of measuring changes in the scattered atom flux.
Further insights into collision dynamics and contrast generation are obtained from incidence energy (\ei{}) dependent measurements, using both He (light) and Kr (heavy) atoms as probes. 

\begin{figure*}[t]
	\centering
	\includegraphics[width = 1\linewidth, draft=false]{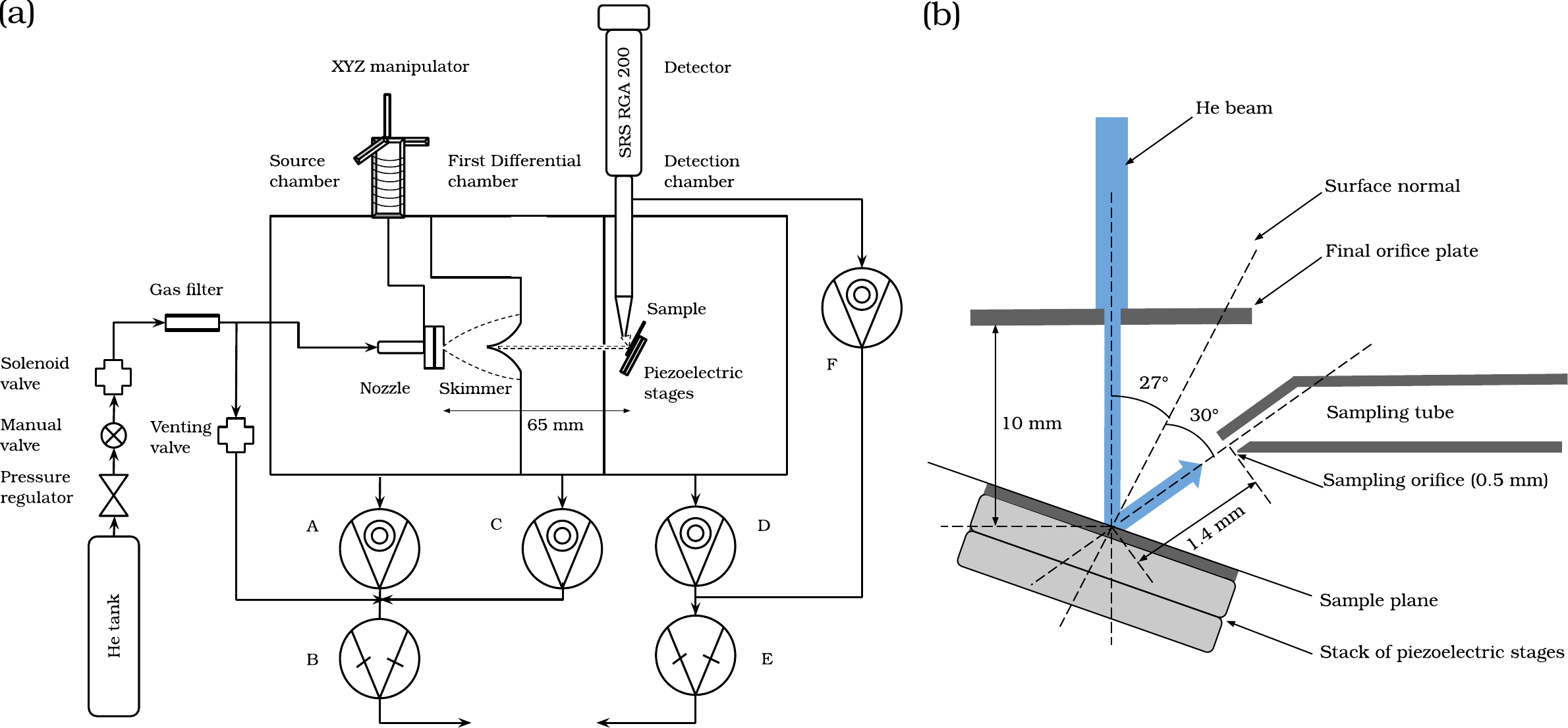}
	\caption{(a) Block diagram of our experimental setup for NAM. He gas at stagnation pressure (typically 6\,bar) is allowed to expand supersonically using a $\mathrm{20\,\mu m}$ nozzle in the source chamber.
	Gas is passed through a $\mathrm{0.5\,\mu m}$ filter to avoid any clogging. Nozzle position can be optimized using an XYZ manipulator. Atomic beam is collimated by a $\mathrm{200\,\mu m}$ opening skimmer and a $\mathrm{20\,\mu m}$ aperture before reaching the sample. Scattered atoms from the target are collected by a sampling aperture connected to a mass spectrometer (flux detection).
	(b) A detailed view of the scattering geometry used in our experiments.
	Incident He beam and sampling aperture are at an angle of $\rm{27^\circ}$ and $\rm{30^\circ}$ with respect to the surface normal, respectively.
	Collection angle of the sampling orifice spans a range of approximately $\rm{20^\circ}$. Sample holder is attached to a pair of piezo-driven stages which enables to perform a raster scan in order to obtain a map of partial pressure vs sample position, leading to NAM images.}
	\label{fig:expt_schematic}
\end{figure*}

\section{\label{sec:level1}Methods}

\subsection{\label{sec:level2}Instrument setup}

Our NAM apparatus produces a collimated atomic beam using a series of pinholes, similar to the designs reported previously \cite{witham, barr_RSI}.
This beam scatters from the sample surface mounted on a movable platform comprising two piezo-driven stages (ECSx3030/NUM(+), Attocube). 
Scattered atoms are detected as a function of sample position by a mass spectrometer (RGA200, Stanford Research Systems) arranged in a flux detection mode.
Figure \ref{fig:expt_schematic}a shows a block diagram of the experimental setup comprising all essential elements from the atomic beam source to the detector. These stages are arranged sequentially as: source, differential, scattering and stagnation detector. 
These are pumped by turbo molecular pumps (denoted by A, C, D and F) with nominal pumping speeds of $\mathrm{1200\,l/s}$ (HiPace 1200, Pfeiffer Vacuum), $\mathrm{300\,l/s}$ (Twistorr 305, Leybold), $\mathrm{300\,l/s}$ (HiPace 300, Pfeiffer Vacuum) and $\mathrm{80\,l/s}$ (HiPace 80, Pfeiffer Vacuum), respectively.
A $\rm{35\,m^3/hr}$ rotary vane pump (Duo 35, Pfeiffer Vacuum) was used to back the source and differential chambers respectively. Scattering and detection stages were backed by an $\rm{11\,m^3/hr}$ rotary vane pump (Duo 11, Pfeiffer Vacuum). 

Atomic beams of He and Kr were produced by a continuous nozzle as the source, built by sandwiching a circular metal film (Lenox Laser Inc.) with an orifice diameter of $\mathrm{20\,\mu m}$ (thickness $\mathrm{30\,\mu m}$) between two metal plates. The beam is extracted into a differentially pumped chamber using a skimmer (Beam Dynamics, Model 2) with orifice diameter of $\mathrm{200\,\mu m}$.
Another metal film having a $\rm{20\,\mu m}$ diameter orifice was used to collimate the beam before entering into the scattering chamber.
Source to sample distance was approximately 65\,mm. 
Incident beam cross-sectional diameters were measured using a knife-edge scan on the sample plane, approximately 10\,mm away from the final aperture. Cross-sectional diameters (full width at half maximum, FWHM) ranged from 27 to 33\,$\mu$m (table \ref{table:incbeam}), corresponding to a narrow angular divergence of $\sim$ 1\,mrad (see appendix \ref{appendix:incbeam_width}). 

\begin{table}[ht]
	\caption{Incident kinetic energy (\ei{}) of He and Kr in meV and their cross-sectional diameters in $\mu$m (FWHM) on the sample plane for different atomic beams used in this work.
	\ei{} values were calculated using equation \ref{equation:v_term} and beam diameters were estimated from knife-edge scan measurements (appendix A). Uncertainties correspond to standard error obtained from fitting. The He beam at \ei{} = 6 meV was not used in the present experiments due to its low incident flux.} 
	
	\begin{tabular}{p{2.9cm} p{1.2cm} p{1.2cm} p{2.7cm} } % centered columns (4 columns)
		\hline\hline %inserts double horizontal lines
		\rule{0pt}{3ex} 
		Composition & \ei(He) & \ei(Kr) & Cross-section dia, FWHM, (He, Kr) \\ %[0.8ex] % inserts table
		%heading
		\hline % inserts single horizontal line
		100\% He & 65 &  - &  26.9$\pm$0.5, \quad - \quad \\ % inserting body of the table
		10\% Kr + 90\% He & 22 & 453 &  30.8$\pm$1.2, 28.9$\pm$0.7\\
		20\% Kr + 80\% He & 13 & 272 &  29.5$\pm$2.2, 30.4$\pm$0.8\\
		30\% Kr + 70\% He & 9 & 194  & 28.8$\pm$2.6, 29.9$\pm$0.8\\
		50\% Kr + 50\% He & 6 & 124 & \quad \quad - \quad \ , 33.4$\pm$1.2 \\ [0.1ex] % [1ex] adds vertical space
		\hline %inserts single line
	\end{tabular}
	\label{table:incbeam}
\end{table}

Typical steady state pressures in the source, differential stages were $\rm{2\times10^{-8}}$ \,mbar, $\mathrm{1\times10^{-8}}$\,mbar with beam off and $\rm{1\times10^{-4}}$\,mbar, $\rm{1\times10^{-6}}$\,mbar with beam on, respectively (pure He, backing pressure 6\,bar).
Pressure increase in the scattering chamber with beam on (measured with an ionization gauge) remained below our detection limit with the pressure being $\rm{1\times10^{-7}}$\,mbar. 
Background He partial pressure (beam off) detected was typically around $\rm{1.25\times 10^{-12}\,mbar}$. 
With the beam on, detected He partial pressure was $\rm{1.55\times 10^{-12}\,mbar}$ (target sample removed).
Considering this change, we estimate that a total of 2.2$\times$10$^9$\,atoms/sec are incident on the target surface corresponding to a flux of 
7$\times$10$^{14}$\,atoms/sec/str (appendix \ref{appendix:incbeam_flux}).

For varying \ei{}, seeded beams of Helium and Krypton with different compositions were employed.
These were prepared by mixing the constituent gases in the appropriate ratio of partial pressures in a stainless steel reservoir and care was taken to allow the gases to mix for at least 6-8 hours prior to any measurement. 
\ei{} of He and Kr ranged from 9 to 65\,meV and 124 to 453\,meV, respectively (table \ref{table:incbeam}). Terminal velocity ($v_{\infty}$) of the gas mixture was estimated using the following relation \cite{scoles1988atomic}:
\begin{equation}
{v_{\infty} = \sqrt{\dfrac{2R}{\langle m \rangle}\left(\dfrac{\gamma}{\gamma - 1}\right)T_0}}
\label{equation:v_term}
\end{equation}
Where $\langle m \rangle$ is atomic mass of the gas (weighted average mass in case of mixtures), $R$ is the universal gas constant, $\gamma$ is the ratio of specific heats and $T_0$ is the stagnation temperature.
\ei{} of He or Kr in a given gas mixture was calculated as $ \frac{1}{2} \ m_i v_{\infty}^2$, where $m_i$ corresponds to the mass of i\textsuperscript{th} component (He or Kr).

Arrangement of incident beam, sample positioning and  detection is depicted in figure \ref{fig:expt_schematic}b.
Sampling aperture is placed approximately along the specular direction with an incident and detection angle of $\mathrm{(27 \pm 1.5) ^{\circ}}$ and $\mathrm{(30 \pm 1.5) ^{\circ}}$ with respect to the surface normal, respectively.
Sampling aperture collects the scattered flux over a relatively large span of $\mathrm{20^{\circ}}$ which comprises of specular reflected signal (0\textsuperscript{th} order diffraction), diffuse elastic and inelastic scattering components (see results and discussion for more details).  
A LabView based program was used to acquire real-time data from mass spectrometer and to control the positioning of stages.

\subsection{\label{sec:level2}Sample preparation and characterization}
Chemical vapor deposition (CVD) method \cite{cvd} was used to grow thin films of molybdenum disulfide (\MoS{}) on $\mathrm{300\,nm}$ thick silicon dioxide ($\mathrm{SiO_2}$) grown on a Si(100) substrate (MicroChem Pvt. Ltd). $\rm{4\,mg}$  ultra pure molybdenum trioxide (99.97$\%$, Sigma-Aldrich) and $\rm{300\,mg}$ sulphur powder  (99.99$\%$ pure, Sigma-Aldrich) were used as precursors (see appendix \ref{appendix:CVD} for details). 
Raman spectroscopic characterization of the samples was done using Renishaw inVia confocal Raman microscope with objectives of 5x and 50x (long focal length). The spectrometer has a spectral resolution of $\mathrm{0.5\,cm^{-1}}$ with a focal spot diameter of $\mathrm{2\,\mu m}$. Excitation wavelength of 532\,nm with $\rm{2\,mW}$ power was used. Acquisition time was set to 20 - 50\,sec for a desired signal-to-noise ratio and measurement speed.

\MoS{} thin film samples used in this work were transferred into the vacuum chamber within 48\,hours of preparation.
Scattered He flux and contrast obtained from thin \MoS{} films wrt  \sio{} substrate were observed to remain largely unchanged over a span of 6 days.
For a given sample, care was taken to complete one set of measurements within this time span (see appendix \ref{appendix:sample_vs_time}). 
For durations longer than 12 to 14 days, a systematic decrease in contrast, obtained using He was observed.
This indicates gradual changes in the sample, possibly caused by hydrocarbon, water adsorption and/or chemical degradation.

\begin{figure*}[t]
	\centering
	\includegraphics[width = 0.98\linewidth]{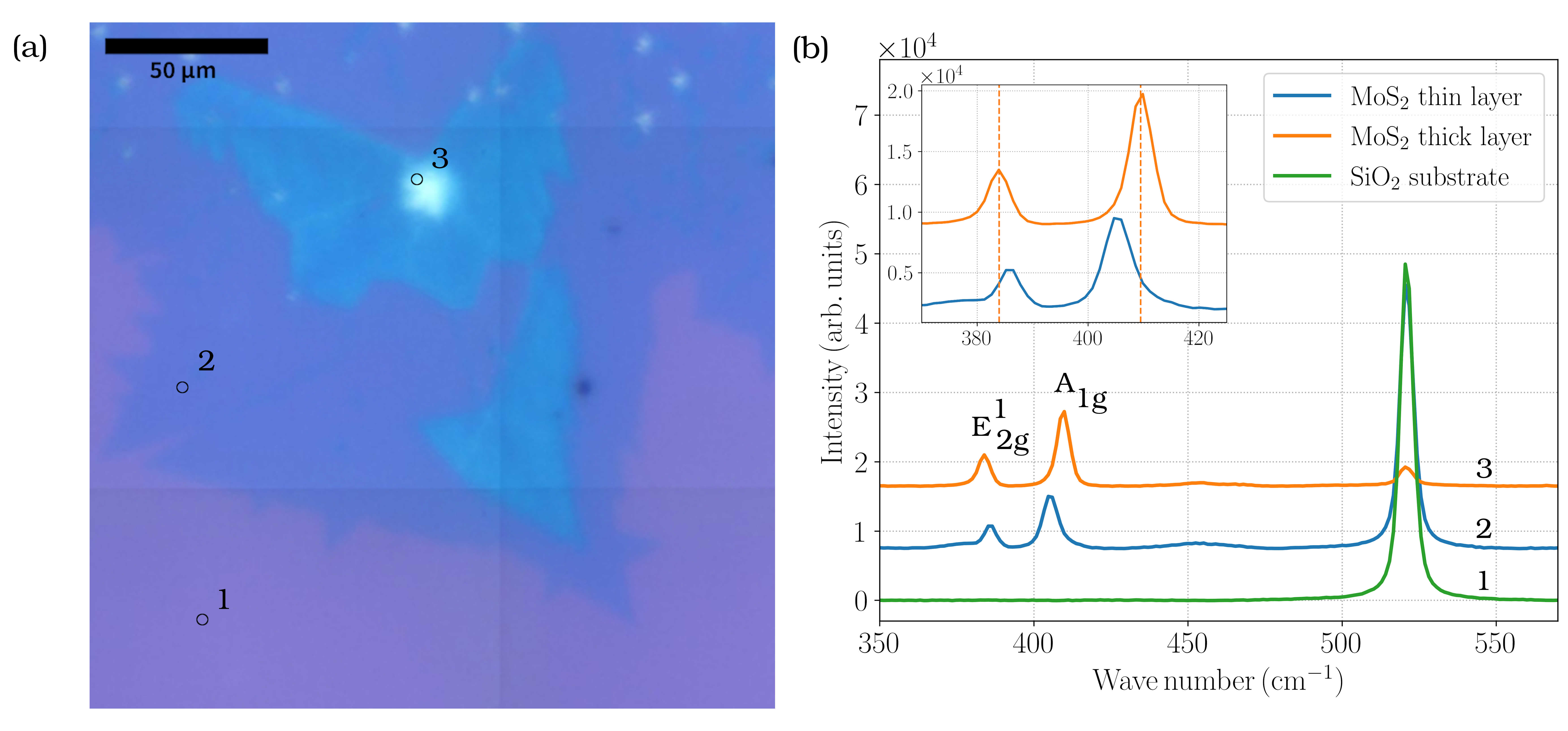}
	\caption{(a) Optical microscopy image of \MoS{} films grown on \sio{} substrate (using white light illumination). Three distinct regions with colors purple (region 1), blue (region 2) and light-blue (region 3) correspond to  \sio{} substrate, thin and thick \MoS{} films, respectively.
		(b) Raman spectra (measured using 2 $\mu$m focal spot diameter) observed at three different regions, measured at positions marked by centers of black circles 1, 2 and 3  shown in panel a. The corresponding spectra (marked as 1, 2 and 3) have been offset along the vertical axis for clarity. In the inset, shifts in the vibrational modes for thin (lower curve) and thick (upper curve) \MoS{} films are clearly visible. Vertical dotted lines correspond to the peak positions of thick (bulk like) \MoS{} films. We assign thin layers to be comprised of one to two monolayers (see text for details).}
	\label{fig:optical_raman}
\end{figure*}

\section{Results and discussions}

\subsection{\label{sec:level2} Characterization of thin films using Optical microscopy and Raman spectroscopy}

Films of \MoS{} on \sio{} substrate, prepared by CVD method, were characterized using optical microscopy and Raman spectroscopy.
An optical image of a small portion of the sample, obtained using white light illumination, is shown in figure \ref{fig:optical_raman}a.
Two distinct features of \MoS{} films are visible, namely, blue and light-blue colored regions on the purple colored substrate. 
Raman spectra measured in the regions of interest (center of black circles in figure \ref{fig:optical_raman}a, 2 $\mu$m diameter illuminated spot) are shown in figure \ref{fig:optical_raman}b.
Spectra obtained for regions 2 and 3 show two additional peaks within (380 to 410)\,\cm{} compared to that for region 1, where a single peak at 521\,\cm{} corresponding to \sio{} \cite{popovic2011raman} is seen, confirming the presence of \MoS{} films. 
The two peaks within (380 to 410)\,\cm{} correspond to in-plane and out-of-plane vibrational modes of \MoS{} with $\rm{E^1_{2g}}$ and $\rm{A_{1g}}$ symmetry \cite{verble}, respectively.

With an increasing number of layers, $\rm{E^1_{2g}}$ and $\rm{A_{1g}}$ modes are known to exhibit a red and blue shift, resulting in a characteristic peak separation as a function of layer thickness \cite{changgu,li2012bulk}.
For samples used in our experiments, peak separation observed in case of thin films was (19 to 20)\,\cm{}, whereas for thick bulk like films, it was (25 to 26)\,\cm{} (see appendix \ref{appendix:Raman_table}). On comparing with previously reported frequency shifts as a function of layer thickness, we conclude that blue colored regions (region 2) correspond to one to two monolayer thickness and light-blue colored regions (region 3) correspond to thicker bulk like layers (greater than six monolayers).

\subsection{\label{sec:level2}NAM measurements and contrast generation mechanisms}

\begin{figure*}[t]
	\includegraphics[width = 1\linewidth]{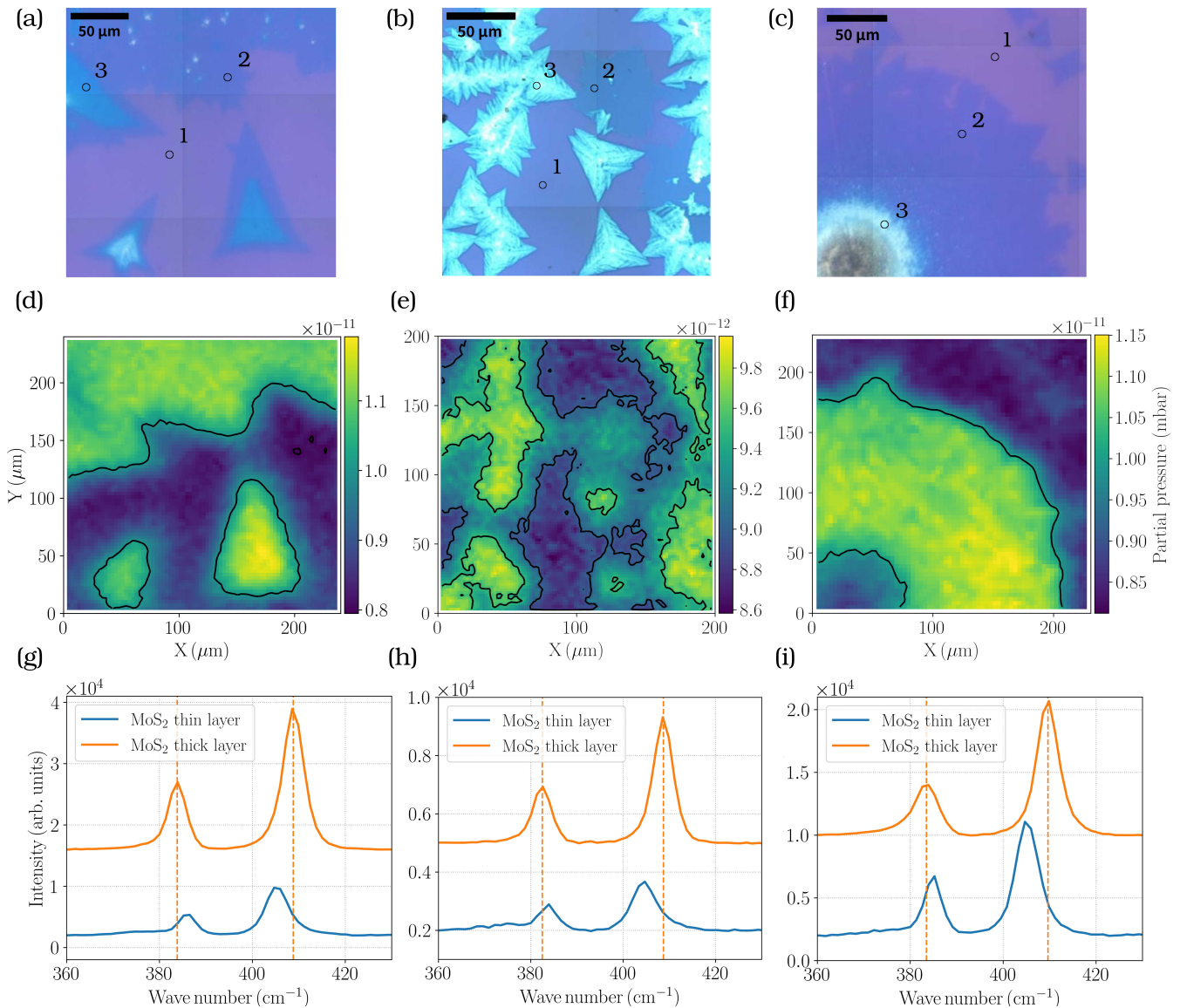}
	\caption{Panels a - c depict optical microscopy images obtained using three independently prepared samples. Panels d - f depict the corresponding NAM images using He, measured with step size of $\mathrm{6\,\mu m}$, $\mathrm{4\,\mu m}$ and $\mathrm{7\,\mu m}$ respectively. The data is linearly interpolated and depending on the features of interest, 2 or 3 iso-partial pressure contour lines are plotted to illustrate the correlated regions.   
	Panels g - i show Raman spectra measured at the points represented by the centers of black circles (region 2 and 3) shown in panels a - c. The upper  and lower curves correspond to thick and thin \MoS{} layers, respectively. Observed peak separations show the presence of a single monolayer (thin) and bulk like \MoS{} (thick) films. A clear one-to-one correspondence among optical and NAM images can be seen. Regions comprising \MoS{} films, as thin as a single monolayer, can be clearly distinguished from the substrate in NAM images.}
	\label{fig:NAM_diffsample}
\end{figure*}

\begin{figure*}[t]
	\includegraphics[width = 1\linewidth]{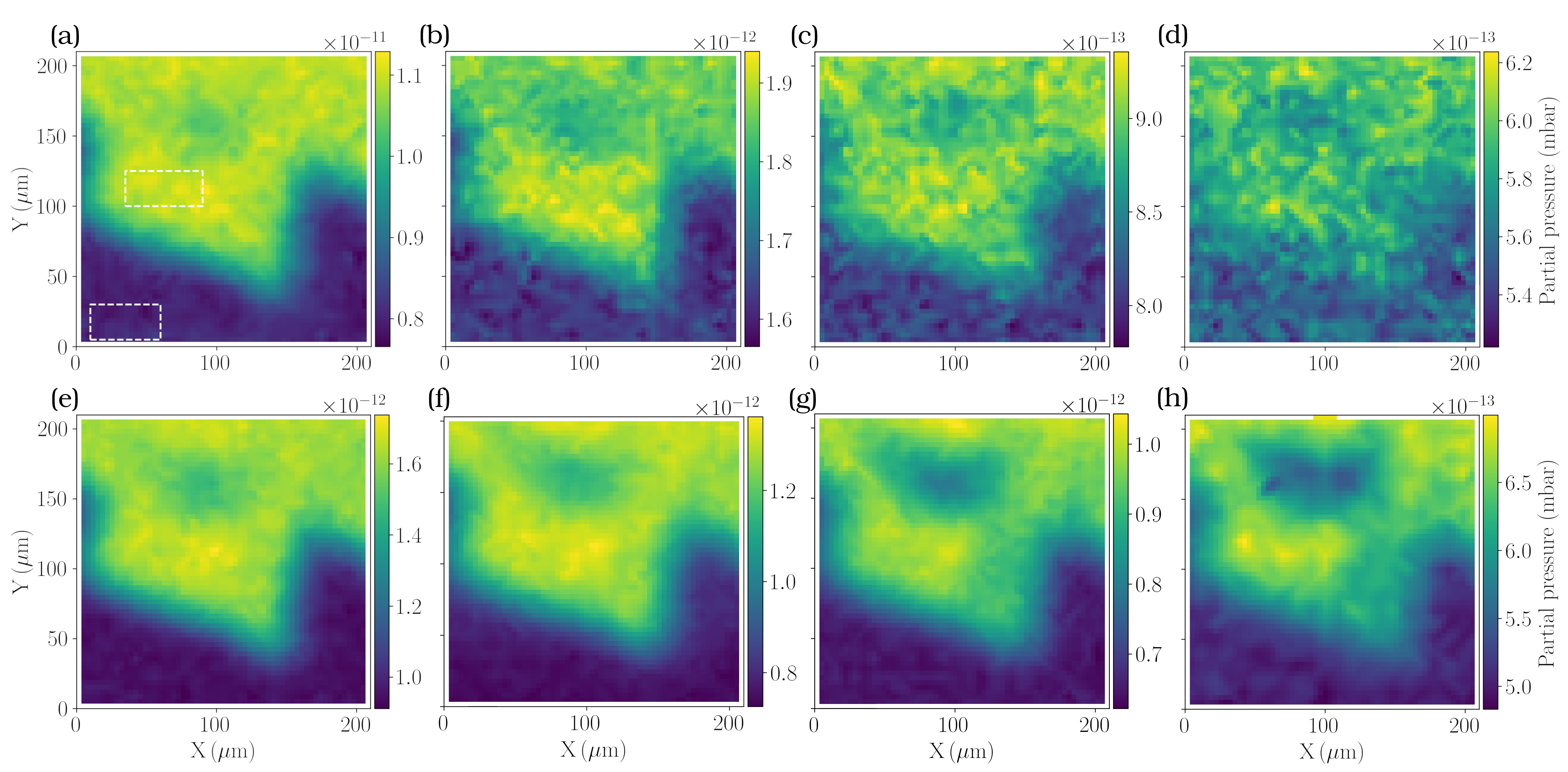}
	\caption{Panels a - d: NAM images obtained using He at different \ei{} for the region shown in figure \ref{fig:optical_raman}a. These measurements were performed at \ei{} of 65, 22, 13 and 9\,meV (left to right). 
	Panels e - h: NAM images obtained using scattered Kr flux with \ei{} of 453, 272, 194 and 124\,meV (left to right). In case of He, a steady decrease in contrast with decreasing \ei{} is observed. Whereas in case of Kr, contrast stays unchanged at higher \ei{} (panels e and f) and decreases at lower energies (panels g and h).} 
	\label{fig:eidep}
\end{figure*}

Optical microscopy images of three independently prepared samples along with Raman spectra are shown in figure \ref{fig:NAM_diffsample}, panels a - c  and panels g - i, respectively.   
As seen previously, in all the samples, bare substrate, thin films and thick films of \MoS{} can be identified using Raman spectroscopy (panels g - h). 
NAM images of the same, obtained by measuring scattered He flux, are depicted in panels d - f.  
It can be seen quite clearly that scattered He flux from  regions covered with \MoS{}, irrespective of layer thickness, is consistently higher by 15-30\% than that from the bare substrate.
A clear one-to-one correspondence among the optical and NAM images can be seen.
The key point here is that, regions with \MoS{} films, as thin as a single monolayer, can be clearly distinguished from the substrate by means of change in scattered He flux captured by our detector.

An important question arising here is regarding the nature of contrast generating mechanism. 
Before delving into interpretation of our experimental observations, we briefly discuss some basic aspects of atom -- surface scattering process, relevant for the question posed above.
An excellent overview of these points is provided in \cite{farias_rev1998, holst_chapter} and references within. 

Scattering of atoms from surfaces can be broadly classified into being elastic and inelastic.
Elastic scattering is a particularly important channel for lighter particles such as He atoms (and H$_2$, D$_2$).
On well-defined pristine surfaces with large flat domains and low defect density, such as polished/cleaved single crystals, He atoms largely undergo elastic scattering.
This results in characteristic diffraction patterns, with the zeroth order peak in specular direction (mirror reflection) being as intense as up to 50 \% relative to the incident beam \cite{comsa_DW_1973, rettner_ArW_1989, farias_focussing2017}.
On the other hand, for heavier atoms such as Ar, Kr, Xe etc., it is well-established that elastic scattering is a relatively minor channel ($\sim$1\%) \cite{rettner_ArW_1989,rettner_ArW1991,rettner_XePt_1991,rettner_ArPt1996}. It can only be distinguished from the relatively larger inelastic scattering fraction by working at low incidence energies and surface temperatures which suppress the inelastic components and thermal broadening, respectively. 

On surfaces with defects such as steps, dislocations, grain boundaries, point defects caused by missing atoms and presence of adsorbates, incident He atoms undergo diffuse elastic scattering.
Cross-sections for diffuse elastic scattering of He atoms due to adsorbates are much larger than their geometric counterparts, with values in the range of few hundred $\mathrm{\AA}^2$.
Further, these cross-sections are known to increase with decreasing \ei{} \cite{comsa_diffuse_1983,farias_rev1998}, leading to a reduction in specular scattered flux at lower \ei{}.

Inelastic scattering mainly arises from energy exchange of atoms with vibrations of surface atoms or adsorbates.
For lighter atoms such as He, this component becomes increasingly important at lower \ei{}, comparable to the lattice vibration energies.
For heavier particles such as Kr atoms, traveling with relatively higher energies as in our experiments, it usually manifests as an energy loss and a relatively broader angular distribution, peaked away slightly from the true specular direction, depending on the momentum exchanged in the collision process \cite{rettner_ArW_1989}.
 
Finally, we would like to point out that in scattering from macroscopically rough surfaces where multi-bounce collisions dominate, elastically scattered flux distributions will also be diffuse. 
For heavier atoms, a large overall energy loss is expected due to multiple inelastic collisions and can possibly lead to trapping.
Several of the above points need to be considered in order to understand the contrast generation and its \ei{} dependence observed in our experiments, both with light He and heavier Kr atoms.

In our experimental setup, the collection angle spans approximately 20$\,^{\circ}$ about the specular reflection direction. Assuming that diffuse scattering (elastic or inelastic) follows a cosine distribution, we estimate that our detector collects about 2\,\% of the total diffuse scattered flux.
On the other hand, the true specular scattered flux having a much  narrow angular distribution ($\ll$ 1$\,^{\circ}$) will be completely captured by our detector. 
Previously reported He scattering studies from clean single crystal \MoS{} surfaces under ultra-high vacuum conditions \cite{farias_MoS2_2019}, show that specular reflected intensity is of the order of 2\% at 300 K.
Since samples in our experiments are far from being under pristine condition, this represents an upper limit to the true specular scattered flux seen by our detector. 
Consequently, it is well possible that the small fraction of diffuse scattered flux and true specular scattered flux reaching our detector are of comparable magnitude. 
This presents several interesting possibilities for contrast generation and its \ei{} dependence which are discussed below.

At the outset, it is useful to estimate the expected contrast for \MoS{} and \sio{} surfaces using He as probe, assuming that variations in specular intensity is the sole contributing factor. 
One should note that this picture is strictly valid only for the case where surfaces are in pristine condition. Since our samples are likely to have defects and adsorbates, this merely represents a limiting scenario.
Specular scattered intensity of He can be estimated using Debye Waller factor \cite{comsa_DW_1973,farias_rev1998}. 
Resulting contrast from two surfaces a and b with atomic masses $M_{a}$ and $M_{b}$, having Debye temperatures of $\Theta_a$ and $\Theta_b $, is governed by the following relation \cite{allison_2003}:
\begin{equation}
C = \tanh \left[ \frac{\alpha}{2} \left(\frac{1}{M_a \Theta_a^2} 
- \frac{1}{M_b \Theta_b^2} \right) \right]
\label{equation:DW_contrast}
\end{equation}
where,
\begin{equation}
\alpha = \frac{24 m (E_{\rm inc} \cos^2 \theta_{\rm inc} + D)T{\rm_s}}{k{\rm_B}}
\label{equation:alpha}
\end{equation}
Here,
$m$ = mass of incident atom (He), \ei{} = incident energy,  $\theta_{\rm inc}$ = incident angle, $D$ = depth of the attractive potential (assumed to be same for both surfaces a and b), $T_{\rm s}$ = surface temperature,  $k{\rm_B}$ = Boltzmann constant.

Unlike surfaces composed of a single kind of atom, in the present case it is not clear a priori what would be the values of $M_a$ and $M_b$. Nevertheless, a qualitative estimate can be made as discussed below. 
For \MoS{}, based on the attenuation of specular He intensity vs surface temperature \cite{farias_MoS2_2019}, we obtain an effective surface mass in the range of 210 - 240\,amu. 
Given that He signal for \sio{} surface is comparatively lower by 10 to 30\,\%, the range of surface mass of \sio{} which satisfies this constraint is around 60\,amu.
This leads to an estimated contrast in the range of 5 to 15\,\% which decreases with \ei{} (appendix \ref{appendix:specular_refl_est}).
For Kr, even from pristine surfaces, inelastic scattering will dominate and the scattered flux will have a broad angular distribution.
At present, width of these angular distributions and its \ei{} dependence are unknown to us and hence an estimate of contrast, as in case of He can not be made.

Another factor that needs to be considered is that of diffraction based contrast generation.
It has been reported in atom scattering based imaging experiments, that in case of He atom scattering from LiF(100) surface \cite{jardine_diffraction}, features corresponding to diffraction can be observed from the regions where sample is locally flat.
Extending this argument one can say that, in general, different surfaces will lead to different diffraction patterns thereby leading to a contrast.
Further, as \ei{} decreases, diffraction peaks are expected to appear at larger angles from the specular direction.
Hence, for a fixed detector geometry placed near specular direction, this can lead to a decreasing signal (and contrast) with decreasing \ei{}. 
Based on the results available for He scattering from \MoS{} surfaces \cite{farias_MoS2_2019} the first order in-plane diffraction peaks from \MoS{} are expected to be approximately at +/- 15 $^{\circ}$ from the specular direction. At lower \ei{} these peaks are expected to move away from specular direction and lowering the signal and contrast, consistent with our observations using He.
A more thorough evaluation of the role of diffraction based contrast generation and its \ei{} dependence can be obtained by measuring angular distributions.

A more realistic scenario for our experiments is that \MoS{} on \sio{} samples have several defects and adsorbates present. 
Under such circumstances, a large fraction  He atoms ($>$ 98\%) are expected to undergo diffuse elastic scattering.
If \MoS{} surfaces have a lower defect/adsorbate density compared to the substrate, it will lead to a relatively lower diffuse elastic scattering component (and larger mirror-like reflection component).
This would be consistent with larger scattered signals observed from \MoS{} surfaces seen in our experiments.
Further, increase in diffuse elastic scattering cross-sections with decreasing \ei{} will result in a loss of signal in the specular direction. 
The exact rate of this change will depend on the specific nature of surface and defects/adsorbates.
In case, this diffuse elastic component from \MoS{} surface increases faster as compared to that on \sio{}, it will result in a decrease in contrast with decreasing \ei{}, as seen in figure \ref{fig:eidep}a -- 4d. 
For Kr atoms, the already broad scattering is not expected to be influenced much with decreasing \ei{}, leading to a little or no dependence.

\begin{figure}[H]
\includegraphics[width = 0.9\linewidth, draft = false]{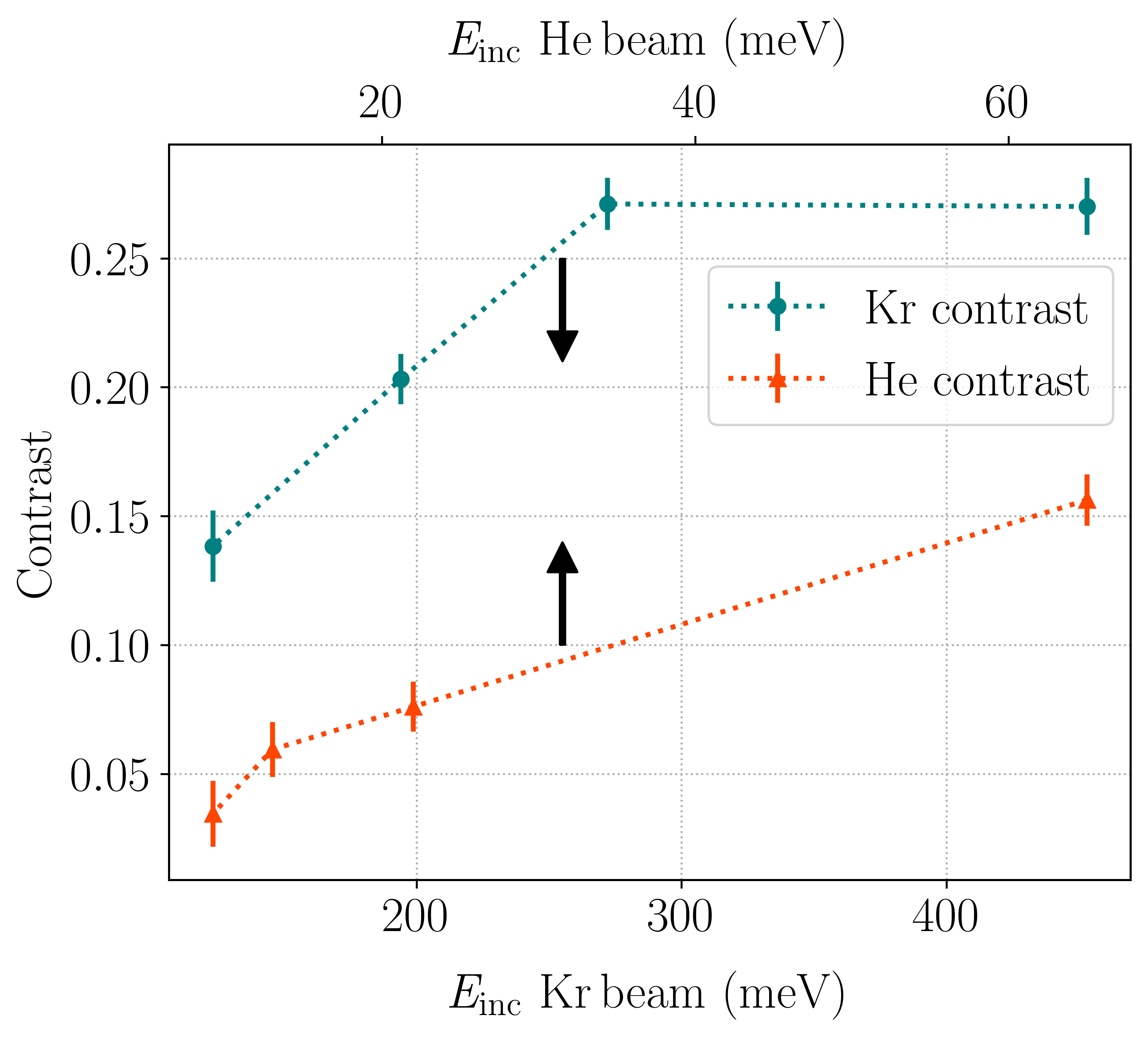}
\caption{Contrast obtained using He (triangles) and Kr (dots) at different \ei{}. These values were obtained from mean signals measured in regions marked by white rectangles in figure \ref{fig:eidep}a, using equation \ref{equation:contrast}. In case of He, contrast decreases monotonically with \ei{}. For Kr, contrast remains unchanged for 453 to 272\,meV but decreases at lower energies. Error bars correspond to one standard deviation. Dashed lines are drawn as a guide to eye for highlighting the trends.}
\label{fig:contrast}
\end{figure}

Different out-of-plane scattering components resulting from inelastic scattering of incident He atoms can lead to changes in specular scattered flux.
Barr and coworkers \cite{barr_NatComm2016} have identified this effect by imaging a series of thin metal films (15 to 40\,nm thickness) on \sio{} surface using He as probe.
Contrast observed in their experiments does not follow the trend expected from surface roughness (measured independently using atomic force microscopy),  but decreases with decreasing incidence energy.
In the present case, for He, a similar trend is observed (figures \ref{fig:eidep} and \ref{fig:contrast}). 
It could well be that changes in the inelastic scattering component of He on \MoS{} and \sio{} lead to an \ei{} dependent contrast in NAM images. 
As mentioned earlier, Kr being relatively heavier and at larger \ei{} than He (see table \ref{table:incbeam}) is expected to undergo largely classical scattering \cite{rettner_ArW_1989,andersson_heavyatom_2002,pollak_heavyatom_2011}.
Under these conditions, scattered flux of Kr will be relatively insensitive to surface lattice vibrations, leading to more or less unchanged contrast as observed in 450 to 270 meV region.
The decrease in contrast at lower energies could possibly arise from trapping and desorption behavior, where Kr atoms leave the surface with a very broad angular distribution leading to a lower signal seen by our detector.

Finally, another point that needs to be looked into is that of change in surface roughness of \MoS{} vs \sio{}.
Although, for the current samples a direct surface roughness measurement is not available with us, we estimate this based on previously prepared samples using the same experimental setup and methodology.
Measurements over regions with areas up to 25 $\mu$m$^2$ show that the surface roughness, characterized by RMS of height distribution, is lower for \MoS{} (1.28 nm - 1.72 nm) than \sio{} (1.99 nm - 3.15 nm) (see appendix \ref{appendix:AFM}). 
Assuming that a similar picture holds true for our samples, this is likely to be important factor in contrast generation, especially when Kr atoms are used as a probe.
The beam of Kr atoms being at higher \ei{} than He, is a more sensitive probe of surface roughness and the increased Kr signal from \MoS{} surface and contrast (compared to He) points towards the same.
It should be noted that the contrast based purely on surface roughness is expected to be independent of \ei{}, hence additional factors as discussed above need to be considered, especially for He atoms.

\section{Concluding Remarks}
Using our recently developed NAM apparatus we demonstrate that thin \MoS{} films up to a single monolayer on a \sio{} substrate can be successfully imaged. NAM images obtained for both He and Kr, exhibit a higher scattered flux from \MoS{} films compared to underlying substrate.
Observations made using Kr atoms point towards the role of decreased surface roughness caused by \MoS{} films, compared to \sio{} substrate, leading to contrast generation. The decrease in contrast at lower \ei{} possibly arises from trapping-desorption of Kr atoms.
On the other hand, in case of He atoms, several factors such as changes in surface roughness, specular component,  diffuse elastic scattering, inelastic scattering, diffraction are possibly playing a role.
At present, based on these results alone, it is difficult to precisely evaluate the contribution of these individual factors.
Nonetheless, a careful measurement of the angular distributions of the scattered flux under different conditions (\ei{} and $T_s$) can provide further insights.

Our apparatus currently does not allow sample/detector rotation and measuring angular distributions, however a relatively large working distance and low angular divergence of incident beam in our experiments can be utilized to set up these experiments in future.
Another interesting possibility is that of using state of the art ion imaging methods  \cite{harding_VMI_2015} to  measure the momentum distribution of scattered atoms. These experiments are expected to provide valuable insights into scattering dynamics and will be the focus of upcoming work in our laboratory.
Nature of \ei{} dependent contrast obtained using Kr atoms suggests an interesting possibility of mapping different surfaces based on differential trapping probabilities. 
Finally, our results illustrate that inherently sensitive nature of atom-surface scattering process can be used in NAM for imaging films as thin as a single monolayer, opening up an important direction in its further development.

\section*{Author contributions}

GB and PRS conceptualized the experiments, designed and tested the experimental NAM setup. GB performed the NAM measurements, wrote  programs for data acquisition and carried out the data analysis. KRS and RS contributed to sample preparation and Raman spectroscopy based characterization along with GB. PN contributed to data analysis and interfacing of piezo stages. GB and PRS prepared the manuscript. All authors discussed the results and provided inputs to the manuscript.
\\

\section*{Competing interests}
The authors declare no competing interests.

\section*{Data Availability}
All relevant data related to the current study are available from
the corresponding author upon reasonable request.

\section*{Acknowledgements}
This work was partly supported by intramural funds at TIFR Hyderabad from the Department of Atomic Energy and Scientific and Engineering Research Board, Department of Science and Technology (grant numbers: CRG/2020/003877 and ECR/2018/001127). 
We thank T. N. Narayanan for insightful discussions regarding sample preparation and characterization, Janmay Jay for help in sample preparation, M. Krishnamurthy for providing vacuum manipulator and pump, Rakesh Mudike for fabricating sampling aperture and sample holder and Saurabh Kumar Singh for helping in Raman spectroscopy measurements. This work was conceived and executed during the ongoing COVID-19 pandemic. We thank the institute staff for their efforts to keep the facilities up and running that enabled this work to be carried out.

\newpage
\onecolumngrid

\appendix
\section{Incident beam width measurement}
\label{appendix:incbeam_width}

Beam width estimation corresponding to that shown in Table 1 of the main manuscript.

\subsection*{He beam:}
\begin{figure*}[h!]
	\centering
	\includegraphics[width = 1\linewidth,trim=5cm 0.8cm 5cm 2.5cm, clip=true]{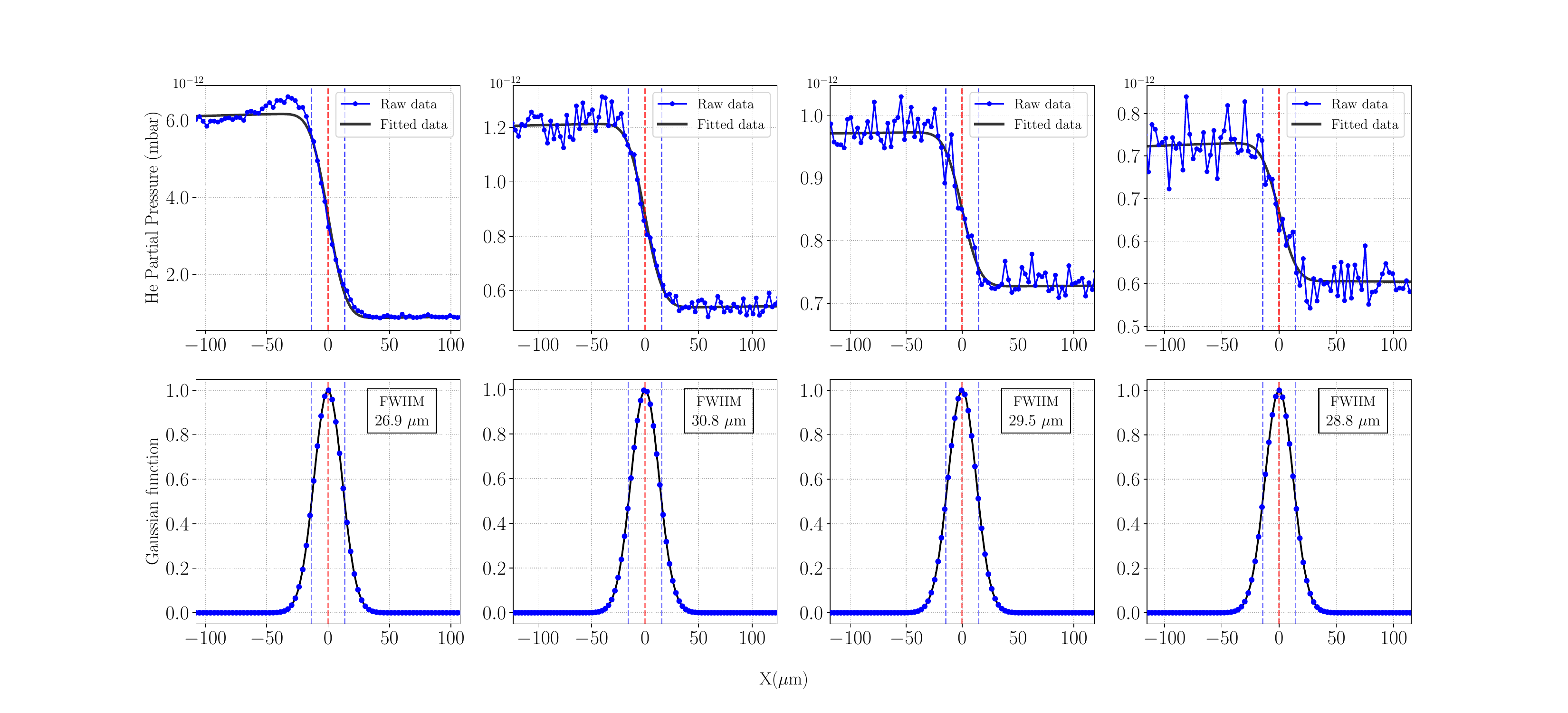}
	\caption{Estimation of the incident beam (He) width on the sample plane using knife-edge scan method.  First row (left to right), beam widths of He in the following mixtures: Pure He, 10\% Kr + 90\% He, 20\% Kr + 80\% He, 30\% Kr + 70\% He, respectively. 
	Blue curve shows He signal observed as a function of position of the knife-edge (razor blade) with step size of $\rm{3\,\mu m}$. Black curve shows the best fit, using a model based on a step function convoluted with a Gaussian function, with three fit parameters namely, amplitude, center and characteristic width. The origin of x-axis is shifted to center of the beam.
	Second row (left to right) show the modeled beam profile with the parameters obtained by fitting. }
	\label{fig:He_beam_width}
\end{figure*}

\clearpage

\subsection*{Kr beam:}

\begin{figure*}[h!]
	\centering
	\includegraphics[width = 1\linewidth, trim=5cm 0.8cm 5cm 2.5cm, clip=true]{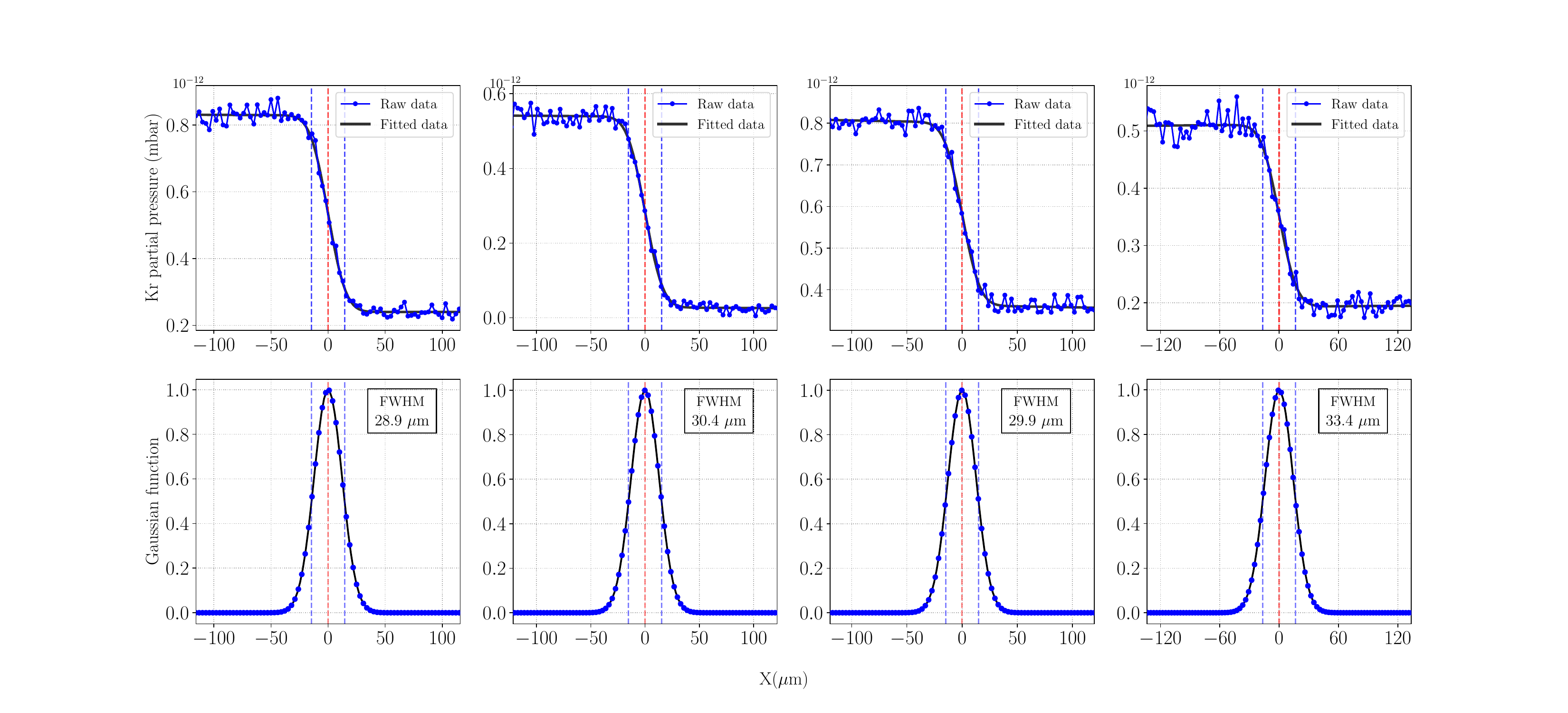}
	\vspace{-0.6cm}
	\caption{Estimation of the incident beam (Kr) width on the sample plane using knife-edge scan method. First row (left to right), beam widths of Kr in the following mixtures: 10\% Kr + 90\% He, 20\% Kr + 80\% He, 30\% Kr + 70\% He and 50\% Kr + 50\% He, respectively. 
	Blue curve shows Kr signal observed as a function of position of the knife-edge (razor blade) with step size of $\rm{3\,\mu m}$. Black curve shows the best fit, using a model based on a step function convoluted with a Gaussian function, with three fit parameters namely, amplitude, center and characteristic width. The origin of  x-axis is shifted to of the beam.
	Second row (left to right) show the modeled beam profile with the parameters obtained by fitting.} 
	\label{fig:Kr_beam_width}
\end{figure*}

\clearpage

\section{Incident beam flux estimation}
\label{appendix:incbeam_flux}

For the final collimation stage, a pinhole of 20 $\mu$m diameter has been used. Considering the change in  partial pressure of gas when beam is on and off, we estimate that $\rm{2.2\times10^{9}}$ atoms/sec are incident on the target surface at a backing pressure of 6 bar. The estimation is shown below:\\
In transparent region (i.e. no sample in front of final aperture):\\
\begin{equation*}
\mathrm{When\,\,He\,\,beam\,\,is\,\,off\,;\,Observed\, partial\, pressure = 1.25\times10^{-12}\,mbar}
\end{equation*}
\begin{equation*}
\mathrm{When\,\,He\,\,beam\,\,is\,\,on\,;\,Observed\, partial\, pressure = 1.55\times10^{-12}\,mbar}
\end{equation*}
\begin{eqnarray*}
	\mathrm{Throughput\,in, Q_{in}} &=& \mathrm{P_{steady state}\times Pumping\,speed} \\
	&=& (1.55\times 10^{-12} - 1.25\times 10^{-12})\times 300 \left(\mathrm{\dfrac{mbar\times l}{sec}}\right) = 9\times 10^{-11} \mathrm{{mbar\times l}\times \dfrac{1}{sec}}
\end{eqnarray*}

The chamber volume, V is diluted with number of atoms, $\mathrm{n_{C}}$ corresponding to above pressure difference. $\mathrm{n_{c}}$ is equivalent to number of atoms incident on the surface. Simply using ideal gas equation:
\begin{eqnarray*}
	%\mathrm{PV} &=& \mathrm{n_{C}RT}\\
	\Rightarrow 9\times 10^{-11}\left(\mathrm{\dfrac{mbar\times l}{sec}}\right) &=& n_{c} \times 8.314\times 10^{-2}\,\mathrm{\left(\dfrac{l\times bar}{K\,moles}\right) \times 300K}\\
	n_c&=& \mathrm{\dfrac{9\times 10^{-14}}{8.314\times 10^{-2}\times 300}\, moles/sec}\\
	&=& \mathrm{3.60\times 10^{-15} moles/sec}\\
	&=& \mathrm{3.60\times 10^{-15}\times 6.023\times 10^{23}\,atoms/sec}\\
	&=& \mathrm{2.2\times 10^{9}\,atoms/sec}\\
	\rm{Incident \ flux} &=& \mathrm{7\times 10^{14} atoms/sec/str}
\end{eqnarray*}

It should be noted that this is a lower limit estimate given that we are measuring the pressure changes using a small, differentially pumped sampling aperture.

%\clearpage

\section{Details of sample preparation using CVD}
\label{appendix:CVD}

Furnace based CVD system, incorporating two temperature zones and a quartz tube where the precursors were placed, was used.
$\mathrm{N_2}$ was used as the carrier gas  with a flow rate of 200\,SCCM during the entire process. This also provides a chemically inert environment.  
Low temperature zone, containing sulphur powder, was maintained at $\mathrm{200^{\circ} C}$. $\mathrm{MoO_3}$, placed in an alumina crucible, was kept in the high temperature zone beside \sio{} substrate at  $\mathrm{650^{\circ} C}$. It took around 30\,min for high temperature zone to reach $\mathrm{650^{\circ} C}$. Actual growth process took around 15\,minutes. To arrest the growth, furnace was opened straight away to cool it down to room temperature. 

\begin{figure}[H]
	\centering
	\includegraphics[width = 0.6\linewidth, clip=true]{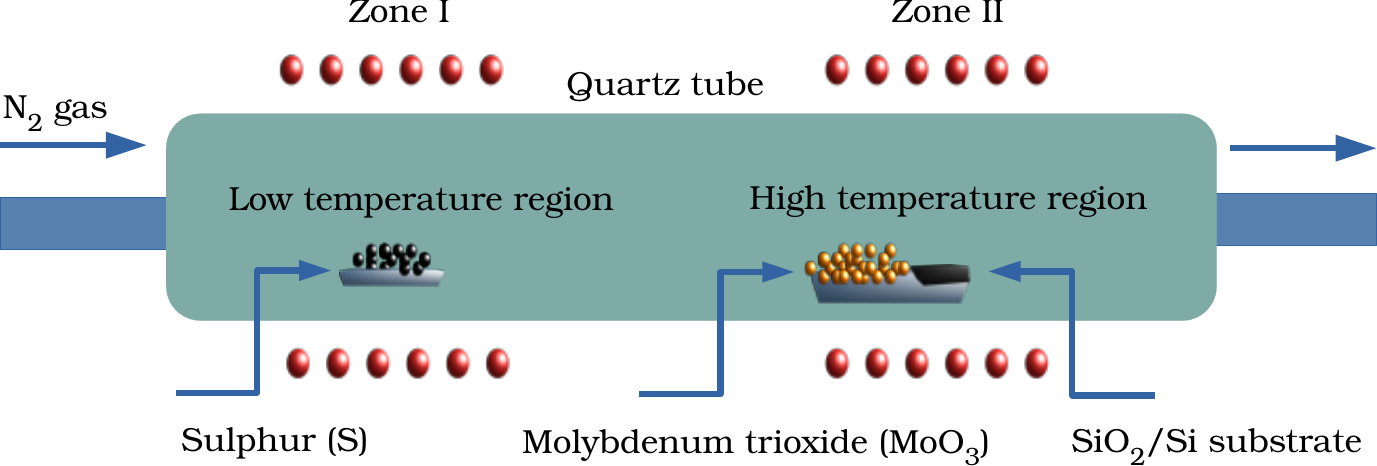}
	\vspace{1cm}
	\caption{A schematic of the setup used for chemical vapor deposition (CVD) having two temperature zones. Precursors are placed at a distance of around $\mathrm{10\,cm}$ from each other. Carrier gas i.e. $\mathrm{N_2}$ is allowed to flow from low to high temperature region for the growth of atomically thin layers of \MoS{}. }
	\label{fig:CVD}
\end{figure}
\vspace{1cm}
\begin{figure}[H]
	\centering
	\includegraphics[width = 0.4\linewidth]{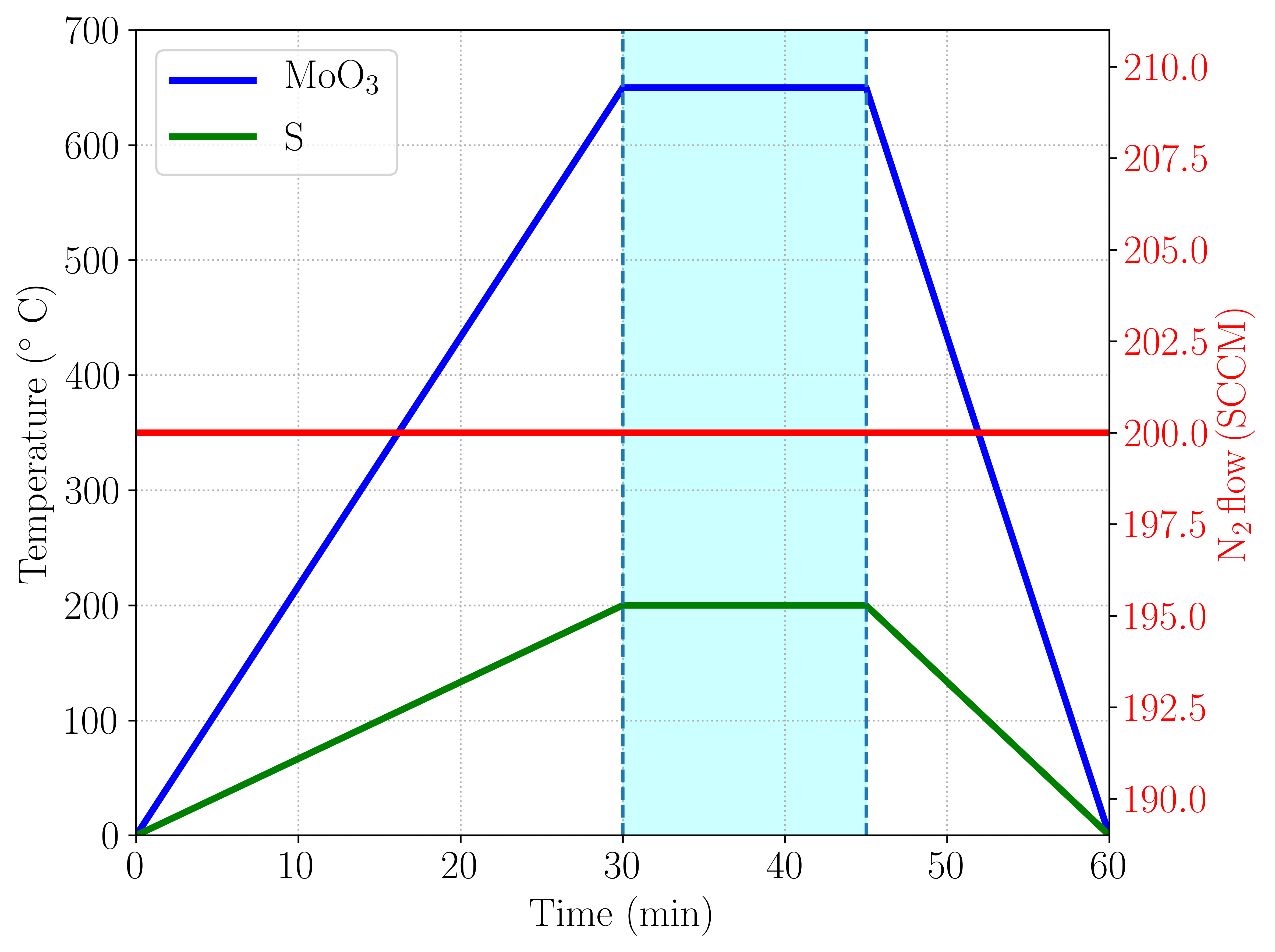}
	\caption{A profile of temperature with time for CVD growth of $\mathrm{MoS_2}$. $\rm{MoO_3}$ placed in high temperature zone takes around 30 min to reach 650$^\circ$ and sulphur powder placed in low temperature zone reaches to 250$^\circ$ in the same time. The growth process takes approximately 15 minutes. These are  optimized temperatures based on the vapor pressure of precursors. }
\end{figure}

%\clearpage

\section{Sample characterization with respect to time}
\label{appendix:sample_vs_time}

Contrast obtained for \MoS{} vs \sio{} using He (pure He beam, \ei{} = 65 meV), measured over a span of six days. Contrast observed for thin \MoS{} films, corresponding to regions A$_1$, A$_2$ and B$_1$, B$_2$, remain largely unchanged over this time span. On the other hand, the contrast measured at different \ei{} (from regions A$_1$, A$_2$) during the same time span shows a much larger change, clearly decreasing with \ei{}. These measurements correspond to the same region of sample as shown in figure \ref{fig:eidep} of the main manuscript.

\begin{figure}[H]
	\centering
	\includegraphics[width = 0.8\linewidth, trim=0cm 0cm 0cm 0cm, clip=true]{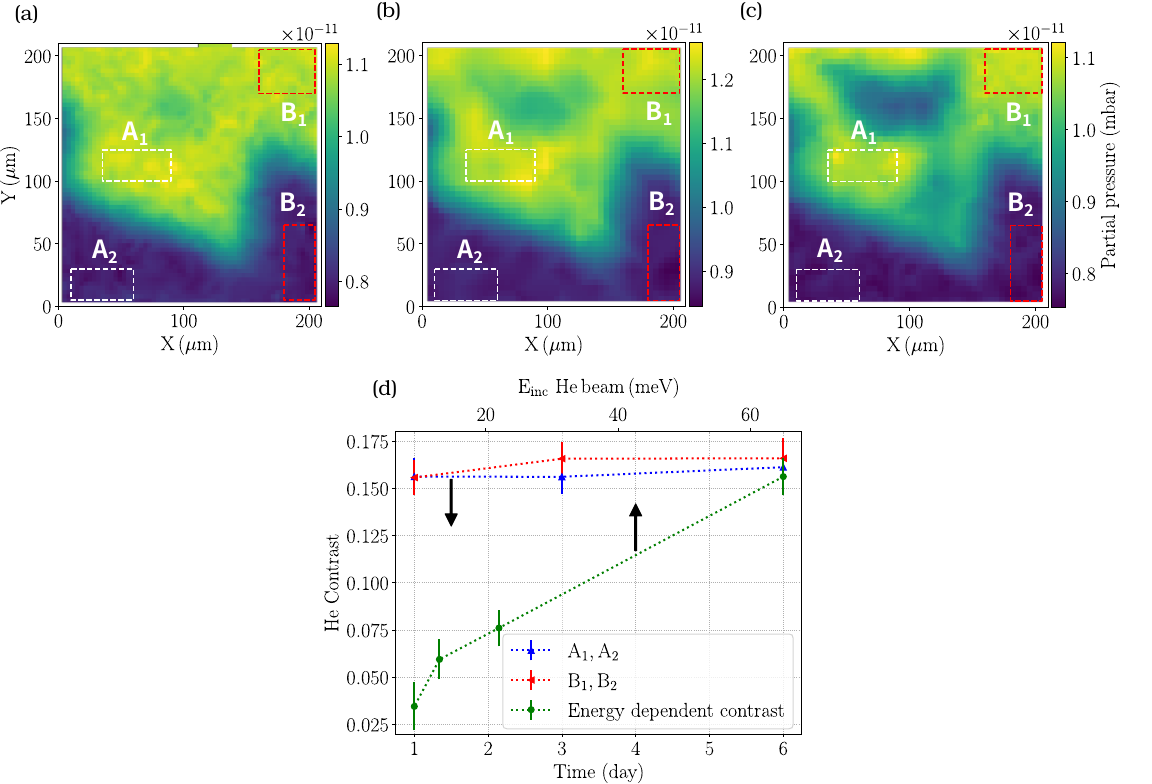}
	\caption{NAM images and contrast measured using He  over a time span of several days. Panels a - c show NAM images (pure He beam at \ei = 65 meV) measured at day 1, 3 and 6 after the sample placed in the vacuum chamber. Panel d, blue and red points show the contrast evaluated from the above images. For comparison, contrast vs \ei{} measured during the same time span is also shown (green points). 
	Contrast is calculated using the signals obtained from regions marked by rectangles and labelled as $\rm{A_1, A_2}$ and $\rm{B_1, B_2}$ in panels a - c. Here A$_1$, B$_1$ and A$_2$, B$_2$ correspond to thin \MoS{} films and substrate, respectively. Dashed lines are drawn as a guide to eye for highlighting the trends. } 
\end{figure}

\clearpage

\section{List of peak positions in Raman spectra}
\label{appendix:Raman_table}

Following table shows the peak positions observed in the Raman spectra corresponding to  $\rm{E^1_{2g}} $ and $\rm{A_{1g}}$ modes in $\rm{MoS_2}$ and that for substrate SiO$_2$. Figure numbers in this table correspond to that in the main manuscript.
\vspace{0.7cm}

\begin{tabular}{ p{3.5cm} p{3cm} p{3cm} p{3cm} p{3cm}  }
	\hline
	\centering
	\textbf{Figure \ref{fig:optical_raman}} & \textbf{$\rm{E_{2g}^1\,(cm^{-1})}$} &\textbf{$\rm{A_{1g}\,(cm^{-1})}$} &\textbf{$\rm{(E_{2g}^1}$} - \textbf{$\rm{A_{1g})\,cm^{-1}}$} &\textbf{$\rm{SiO_2\,peak\,(cm^{-1})}$}\\
	\hline
	$\rm{MoS_2}$ thin layer   &385.80   &405.31  &19.51  &520.86\\
	$\rm{MoS_2}$ thick layer  &383.97   &409.48  &25.51  &520.69\\
	Bare substrate            &         &        &       &520.86\\
	\hline
	\centering
	\textbf{Figure \ref{fig:NAM_diffsample}a} & \textbf{} &\textbf{} &\textbf{} &\textbf{}\\
	\hline
	$\rm{MoS_2}$ thin layer   &385.97   &405.15  &19.17  &520.86\\
	$\rm{MoS_2}$ thick layer  &383.80   &408.81  &25.01  &520.86\\
	Bare substrate            &         &        &       &520.86\\
	\hline
	\centering
	\textbf{Figure \ref{fig:NAM_diffsample}b} & \textbf{} &\textbf{} &\textbf{} &\textbf{}\\
	\hline
	$\rm{MoS_2}$ thin layer   &383.82   &404.46  &20.64  &520.47\\
	$\rm{MoS_2}$ thick layer  &382.49   &408.73  &26.24  &520.74\\
	Bare substrate            &         &        &       &520.47\\
	\hline
	\centering
	\textbf{Figure \ref{fig:NAM_diffsample}c} & \textbf{} &\textbf{} &\textbf{} &\textbf{}\\
	\hline
	$\rm{MoS_2}$ thin layer   &384.97   &404.98  &20.01  &520.86\\
	$\rm{MoS_2}$ thick layer  &383.47   &409.65  &26.18  &520.86\\
	Bare substrate            &         &        &       &520.86\\
	\hline
\end{tabular}

%\clearpage

\section{Estimation of specular reflected intensities considering pristine surfaces}
\label{appendix:specular_refl_est}

Contrast estimation assuming pristine surfaces, where the changes in specular intensity alone is the major factor, can be estimated using Debye Waller factors (using equations \ref{equation:DW_contrast} and \ref{equation:alpha} in the main manuscript).
This requires the masses of surface atoms to be known. 
In the present case of \MoS{} and \sio{} (unlike elemental surfaces comprising single type of atom) we estimate the effective surface mass in the following manner:
The available data of specular scattering vs $T_{\mathrm s}$ has been reported in reference [36].
A fit to this data (two curves) using the surface mass as fit parameter gives the effective mass of \MoS{} surface to be 210 and 240 amu.
For SiO$_2$ such data is not available to the best of our knowledge. However, a reasonable guess can be made based on the constraints available from our results. 

\begin{figure}[H]
	\centering
	\includegraphics[width = 0.8\linewidth, trim=0cm 0cm 0cm 0cm, clip=true]{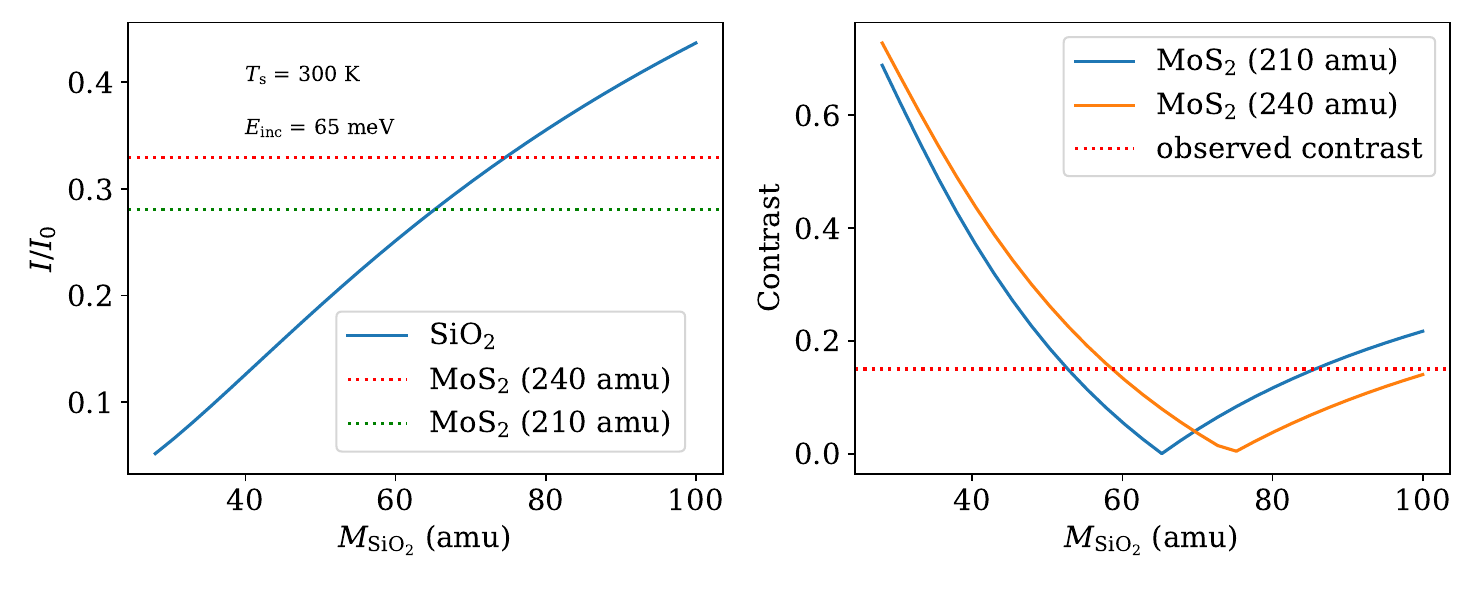}
	\caption{(left) A plot of the specular intensity ($I/I_0$) calculated using Debye waller factor with different surface mass for SiO$_2$. The horizontal dashed lines show the specular intensity calculated for \MoS{} using surface masses 210 and 240 amu. (right) Dependence of the calculated contrast on the surface mass of SiO$_2$. Surface temperature of 300 K and \ei{} of 65 meV is used for these calculations.
		Based on the experimental observations, where the specular reflected signal from \MoS{} is higher than substrate by  approximately 30\% and the contrast about 15 \%, we estimate that effective surface mass of approximately 60 amu seems appropriate.} 
	\label{fig:meff}
\end{figure}

Our experiments show that the scattered signal from \MoS{} surfaces is higher than \sio{} by approximately 30\%, and the typical contrast being 15 \%, we can narrow down the range of effective mass of SiO$_2$ surface.
Estimations depicted in figure \ref{fig:meff} show that given the above constraints, the surface mass for \sio{} will be in the range of 50 - 60 amu.

\begin{figure}[H]
	\centering
	\includegraphics[width = 0.7\linewidth, trim=0cm 0cm 0cm 0cm, clip=true]{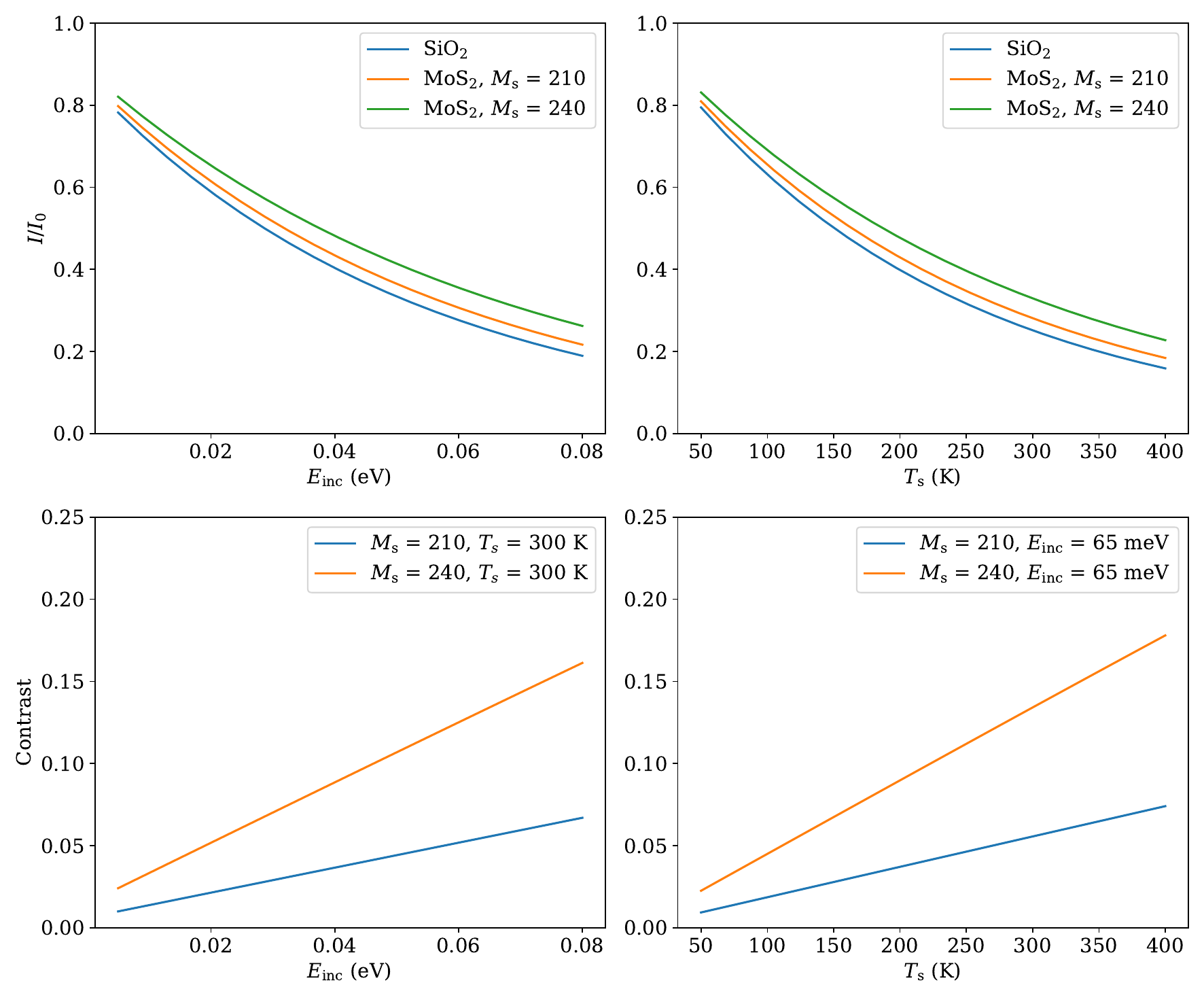}
	\caption{The specular intensities and contrasts are calculated as a function of surface temperature and incidence energy. A surface mass of 60 amu for SiO$_2$ is used (based on the estimation in figure \ref{fig:meff}). (top row) The specular intensity for estimated for \MoS{} (using two different surface masses) and SiO$_2$. Specular intensity decreases with increasing incidence energy and surface temperature. (bottom row) Estimated contrast increases with increasing incidence energy (at surface temperature = 300 K) and also with surface temperature (at incidence energy = 65 meV).}
	\label{fig:dw}
\end{figure}

For the parameters chosen and within the validity of framework where contrasts depend solely on specular scattered signals, contrast is expected to increase at higher \ei{} and also with increasing surface temperature (figure \ref{fig:dw}, bottom row).

\clearpage

\section{Atomic force microscopy (AFM) based roughness estimation}
\label{appendix:AFM}

\begin{figure}[H]
	\centering
	\includegraphics[width = 0.6\linewidth]{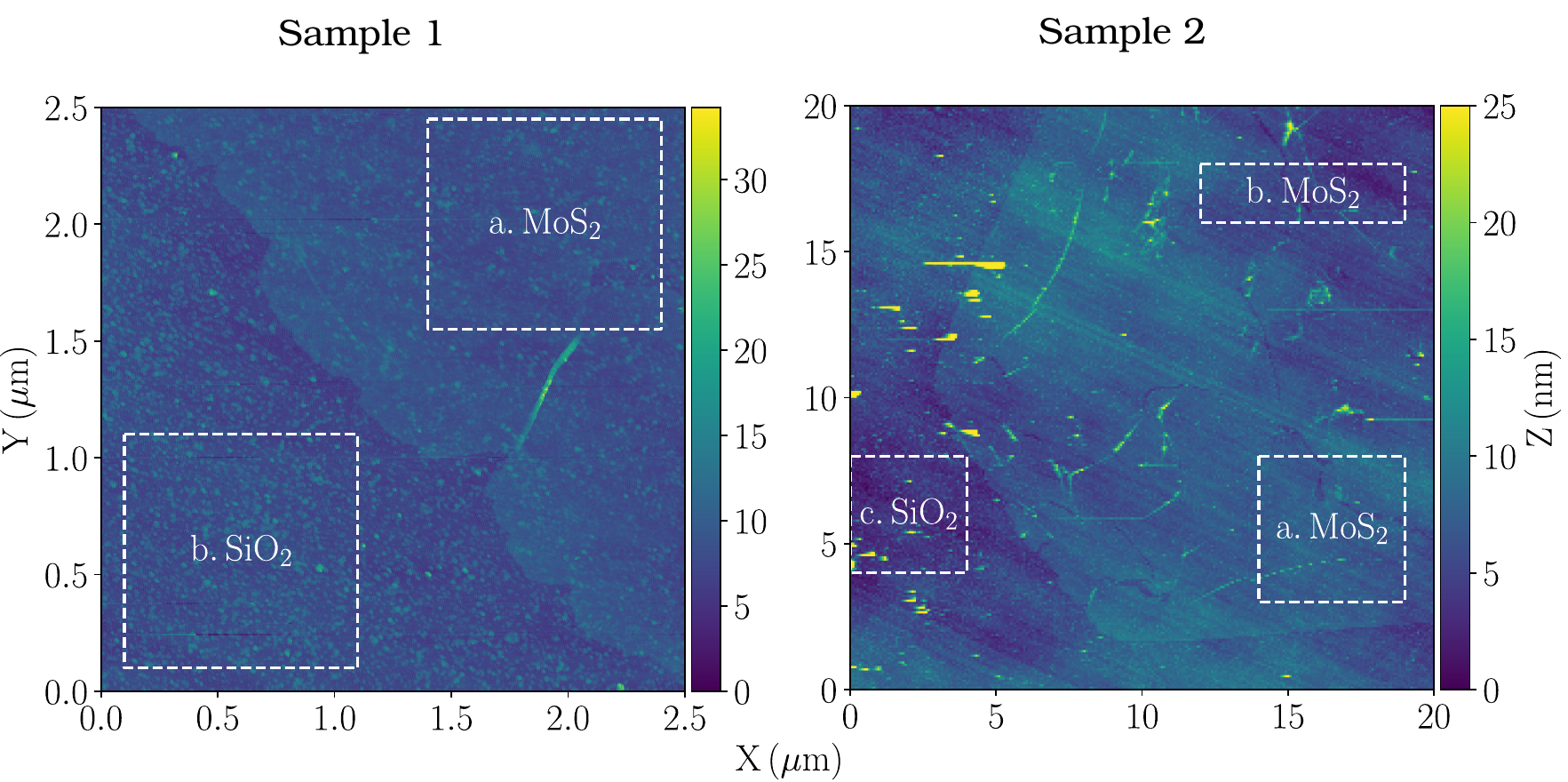}
	\caption{AFM images of two different samples prepared (prepared independently) using the same method and experimental setup (note the different sample sizes). On sample 1, we selected a region of $\rm \sim (1\times 1)\,\mu m$ on $\rm{MoS_2}$ and $\rm{SiO_2}$ as well. Relatively large areas $\rm \sim (5 \times 5)\,\mu m$ were selected on sample 2 to calculate the RMS roughness. These regions are indicated by white dashed rectangles.}
	\label{AFM} 
\end{figure}

\begin{figure}[H]
	\centering
	\includegraphics[width = 0.65\linewidth]{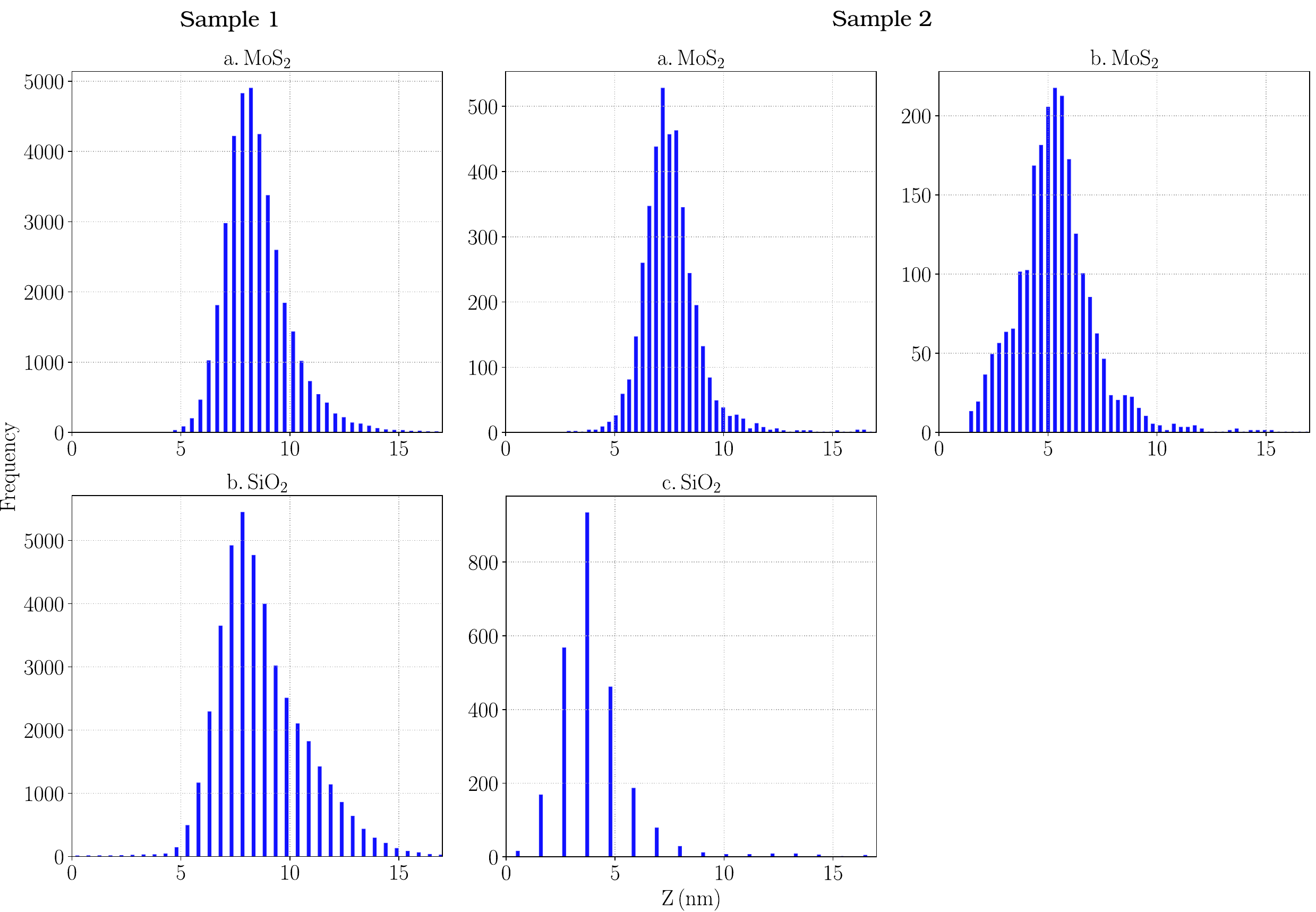}
	\caption{Shows the corresponding histograms of the marked regions in figure \ref{AFM} (above). The X and Y axis in all graphs correspond to the height (Z) in nm and the frequency, respectively.} 
\end{figure}

Following table shows RMS roughness calculated corresponding to $\rm{MoS_2}$ and $\rm{SiO_2}$ regions in two different samples. It shows that surface roughness for $\rm{MoS_2}$ is lower than substrate. These measurements were carried out using the AFM (Dimension Icon, Bruker) at Weizmann Institute of Science, Israel.
\vspace{1.5em}

\centering
\begin{tabular}{ p{4cm} p{3cm} p{4cm}}
	\hline
	\centering
	& \textbf{Regions} & \textbf{RMS roughness (nm)}\\
	\hline
	\textbf{Sample 1}& a. $\rm{MoS_2}$ &1.45\\
	& b. $\rm{SiO_2}$ & 1.99\\
	\hline 
	\textbf{Sample 2} & a. $\rm{MoS_2}$ &1.28\\
	& b. $\rm{MoS_2}$ & 1.72\\
	& c. $\rm{SiO_2}$ &3.15\\
	\hline
\end{tabular}

\clearpage

\twocolumngrid

\bibliographystyle{unsrt}
\bibliography{references.bib}

\end{document}